\documentclass[amsmath,twocolumn,superscriptaddress,showpacs,aps,prl]{revtex4}

\usepackage{bm}
\usepackage{graphics}
\begin{document}
\title{Shot noise in a diffusive F-N-F spin valve}

\author{E. G. Mishchenko}
\affiliation{Lyman Laboratory, Department of Physics, Harvard
University, MA 02138}
\begin{abstract}
Fluctuations of electric current in a spin valve consisting of a
diffusive conductor connected to ferromagnetic leads and operated
in the giant magnetoresistance regime are studied. It is shown
that a new source of fluctuations due to spin-flip scattering
enhances strongly shot noise up to a point where the Fano factor
approaches the full Poissonian value.

\end{abstract}
 \maketitle

Transport in various spintronic devices \cite{ALS} containing
ferromagnet-paramagnet interfaces is attracting a lot of
attention. Considerable experimental and theoretical efforts have
been directed towards the understanding of magnetoresistance, spin
injection, spin accumulation, spin-orbit interaction,
current-induced torque and other fascinating and challenging
effects (the vast and quickly expanding bibliography is far beyond
the scope of this Letter). Advances in technology and sample
fabrication resulting in devices of nanoscale dimensions led the
methods and notions of spintronics to be the natural outgrows and
further developments of the exciting and successful ideas of
mesoscopics.

One of the issues outstanding in mesoscopic physics has been the
phenomenon of the shot noise, i.e.\ current fluctuations in
nonequilibrium conductors \cite{BlB}. In particular, an
experimental confirmation \cite{SMD} of the theoretically
predicted ${1}/{3}$-suppression (compared to the Poissonan value
characteristic for the transmission of independent particles) of
the noise signal in diffusive conductors \cite{BB,Nag} is one of
the milestones in the field. Shot noise in ferromagnet-normal
metal constrictions is also evolving into a subject of much
interest. Current fluctuations in a F-quantum dot-F system in the
Coulomb blockade regime  were considered in Refs.\
\cite{BMM,Bul,LL,SEJ}, noise in a quantum dot in the Kondo regime
analyzed  in Ref.\ \cite{LS}, ballistic beam splitter with
spin-orbit interaction discussed in Ref.\ \cite{EBL}. Dependence
of the shot noise in a diffusive conductor attached to
ferromagnetic reservoirs on the relative angle between the
magnetizations of reservoirs has been studied in Ref.\ \cite{TB}
with the help of the circuit theory \cite{BNB}. However, effects
of a spin-flip scattering on the fluctuations of electric current
in diffusive conductors have been disregarded so far. In the
present Letter we show them to make a profound effect on the shot
noise power.

The universal $1/3$-shot noise in a conventional diffusive
conductor is due to the interplay of the random impurity
scattering and restrictions imposed by the  Fermi statistics. In
the presence of ferromagnetic contacts, however, the spin
degeneracy is lifted with spin-up and spin-down electrons
representing two different subsystems. The number of particles in
each subsystem is not conserved (due to spin-flip scattering)
leading therefore to a new class of fluctuations. The situation
here resembles closely the fluctuations of radiation in random
optical media \cite{Ben}. The absence of particle conservation in
a gas of photons results in the enhancement of photon flux noise
above the Poissonian value (also the result of bunching typical
for bosons). With the notable difference in statistics (Fermi
instead of Bose) the framework of stochastic diffusion equations
\cite{MB,MPB} can be formulated for the fluctuations in disordered
spintronic devices as well.
\begin{figure}[h]
\resizebox{.33\textwidth}{!}{\includegraphics{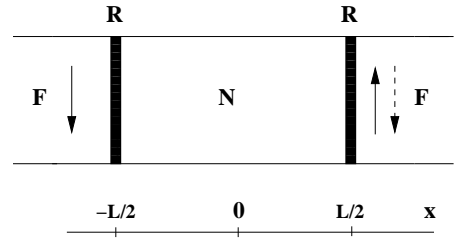}}
\caption{Spin valve consisting of a paramagnetic diffusive
conductor (N) connected to ferromagnetic leads (F) through tunnel
contacts. Conduction bands in the leads are assumed to be
completely spin-polarized. For small amount of spin-flip
scattering the resistance to electric current is large when the
magnetizations are antiparallel ('off'-state of the valve),
compared to the usual metallic resistance for the parallel
configuration ('on'-state).}
\end{figure}

To demonstrate this we discuss the most characteristic example of
a spin valve in the giant magnetoresistance regime, when the
transport across the valve is extremely sensitive to the intensity
of a spin-flip scattering. Namely, we consider a diffusive
paramagnetic conductor (N) sandwiched between two ideal
ferromagnetic (F) leads, Fig.\ 1. 'Ideal' means that electron
distributions inside the leads are not affected by the presence of
the normal region (a typical mesoscopic setup assuming the
conduction and screening in the leads to be more efficient than in
the conductor). In addition, we assume that conduction electrons
are completely polarized inside the ferromagnets, i.e.\ the
population of carriers with a spin direction opposite to that of a
magnet is fully depleted (half-metallic ferromagnets). Therefore,
when the polarizations of the leads are antiparallel, a conduction
electron cannot be transferred across the valve without changing
its spin direction. As a result the resistance of a spin-filter is
very large unless there is a substantial amount of spin-flip
scattering inside the N-region. We assume the F-N interfaces to be
spin-conserving but allow for the finite contact (tunnel)
resistances $R$.

{\it Stochastic diffusion equations.} The electron motion inside
the N-region is diffusive with the mean free path much smaller
than the size of the valve $L$ (but yet much larger than the Fermi
wavelength). At temperatures low enough the inelastic
(electron-phonon, electron-electron) scattering is suppressed
(once the inelastic scattering length exceeds $L$). The electron
distribution is therefore almost isotropic in momentum space and
can be described by the spin and energy-dependent distribution
functions $f_\alpha(x,\epsilon)$, with $\alpha =\pm$ being a  spin
index: $+$ corresponding to spin-up electrons and $-$ to spin-down
electrons.

If the system is driven out of equilibrium (e.g.\ by applying a
voltage bias to the leads),  the distribution function becomes
spatially inhomogeneous resulting in the electric current (we
assume the cross-sectional area of the valve to be equal to
unity),
\begin{equation}
\label{current}
 { j}_\alpha (x,\epsilon,t) = -\frac{\sigma}{e}
\frac{\partial f_\alpha (x,\epsilon)}{\partial x} +{\cal
J}_\alpha(x,\epsilon,t),
\end{equation}
where $\sigma = e^2 \nu D$ is the conductivity in the N-region,
$\nu$ is the density of states per single spin direction and $D$
is the diffusion constant. The last term in Eq.\ (\ref{current})
is the stochastic Langevin source. It has zero expectation value
and a correlator that similarly to the spinless case \cite{Nag} is
determined by the mean value of the electron distribution
function,
\begin{equation*}
\overline{ {\cal J}_\alpha(x,\epsilon,t) {\cal
J}_\beta(x',\epsilon',t')} =2\sigma \delta_{\alpha\beta}\Delta
 \overline f_\alpha (x,\epsilon)[1-\overline f_\alpha (x,\epsilon)],
\end{equation*}
where we have abbreviated $\Delta= \delta(x-x')\delta
(\epsilon-\epsilon')\delta(t-t')$ and assumed no summation over
the repeated indexes. The stochastic source ${\cal J}_\alpha$ is
due to the random independent (i.e. Poissonian) events of
spin-conserving scattering from disorder.

The particle conservation implies a second relation between the
electric current and particle density (hereinafter we drop the
arguments when it could not lead to confusion),
\begin{equation}
\label{second}
 -\frac{\partial f_\alpha}{\partial
t}+\frac{e}{\sigma}\frac{\partial j_\alpha}{\partial x}= \frac{
D}{2L_s^{2}} (f_{-\alpha}-f_\alpha)+ \alpha {\cal L}.
\end{equation}
The first term in the right-hand side accounts for the average
particle flow between states with opposite spins due to spin-flip
scattering (customary in treating spin-dependent diffusion
problems \cite{SKW}). The spin-flip length $L_s$ is assumed to be
much larger than the mean free path but no restrictions as to its
relation to the size of the system $L$ are imposed. The last term
in Eq.\ (\ref{second}) is the Langevin source for the spin-flip
scattering arising from randomness of a spin-flip process. It is
similar to the stochastic terms for the fluctuations of the number
of photons in disordered optical media \cite{MB}. Its second
moment is equal to the mean flow between states with different
spin directions,
\begin{equation}
\label{conserv}
 \overline{ {\cal L}(x,\epsilon,t) {\cal
L}(x',\epsilon',t') } = \frac{D\Delta}{2\nu
L^2_s}\sum_\alpha\overline f_\alpha  (1-\overline{f_{-\alpha}}).
\end{equation}
which utilizes the fact that spin-flip scattering events are
independent and obey Poissonian statistics. In writing Eqs.\
(\ref{second}-\ref{conserv}) we suggested that the spin-flip
scattering is energy-conserving. This assumption is well justified
whenever a typical energy change during a spin-flip is small
compared to the characteristic scale of the electron distribution
(set by the temperature $T$ or external bias $eV$).

 The above equations must be supplemented with appropriate
boundary conditions. We assume that the interface resistances at
the left and right contacts are the same $R$. Since there is no
charge accumulation in the system, the diffusive currents
(\ref{current}) should match the tunneling currents through the
interfaces. In particular, for the antiparallel valve
configuration the boundary conditions read,
\begin{equation}
\label{bound}
\begin{array}{lll} j_-=\frac{1}{eR}[f_L-f_-]+{\cal
I}_L, & j_+=0,& \text{at}~x=-\frac{L}{2}\\
j_+=\frac{1}{eR}[f_--f_R]+{\cal I}_R,& j_-=0, &
\text{at}~x=\frac{L}{2}.
\end{array}
\end{equation}
For the parallel configuration one has to interchange $+$ and $-$
indexes in the second line of Eq.\ (\ref{bound}). The stochastic
sources ${\cal I}_L$ and ${\cal I}_R$ accounting for the
randomness of the electron tunneling through the interfaces have
(at $T=0$) the variance \cite{t},
\begin{eqnarray}
\overline{ {\cal I}_i (\epsilon,t){\cal I}_k(\epsilon',t') }
&=&\delta_{ik}\delta(\epsilon-\epsilon')\delta(t-t') e  \overline
J(\epsilon),
\end{eqnarray}
here $J(\epsilon)=\sum_\alpha j_\alpha(x,\epsilon)$ is the total
current independent of the coordinate $x$, as readily seen from
Eq.\ (\ref{second}). The current at the contacts is due to
electrons with a single spin direction only.

It is convenient to use the particle density and spin density
distributions as well as the corresponding Langevin sources,
\begin{equation*}
 f,f_s =\frac{1}{2} (f_+ \pm f_- ),~~~
{\cal J}, {\cal J}_s =\frac{1}{2}({\cal J}_+ \pm {\cal J}_-).
\end{equation*}
Combining Eqs.\ (\ref{current}) and (\ref{second}) we obtain (in
the stationary regime) the equations for the particle and spin
distribution,
\begin{equation}
\label{equations} \frac{\partial^2 f}{\partial
x^2}=\frac{e}{\sigma} \frac{\partial \cal J}{\partial x}, ~~~~
\frac{\partial^2 f_s}{\partial x^2}= \frac{f_s}{L_s^{2}}+
\frac{\cal L}{D}+ \frac{e}{\sigma} \frac{\partial {\cal
J}_s}{\partial x}.
\end{equation}
Note, that different Langevin terms (${\cal I},{\cal J},{\cal L}$)
are independent and have zero cross-correlators.

{\it Average electric current.} The mean (averaged over time)
solution of the equations (\ref{equations}) with the boundary
conditions (\ref{bound}) is elementary and  yields the
distribution function,
\begin{equation}
\label{distr} \overline f_\alpha (x)
=\frac{f_L+f_R}{2}-\frac{R_s}{2R_0} [f_L-f_R] \left( \frac{x}{L_s}
-\alpha M(x) \right),
\end{equation}
with $R_s=L_s/\sigma$ standing for the characteristic resistance
on a spin-flip length $L_s$. The total resistance $R_0$ and the
function $M(x)$ depend on the magnetization of the leads. For the
antiparallel configuration,
\begin{equation}
R_0=R_N+2R+R_s\coth{s}, ~~~ M(x)=\frac{\cosh{(x/L_s)}}{\sinh{s}},
\end{equation}
while for the parallel configuration,
\begin{equation}
R_0=R_N+2R+R_s\tanh{s}, ~~~ M(x)=\frac{\sinh{(x/L_s)}}{\cosh{s}}.
\end{equation}
Here $s=L/2L_s$ is the dimensionless measure of the amount of
spin-flip scattering in the system, and $R_N=L/2\sigma$ is the
resistance of the normal region.

The total mean electric current calculated with the help of Eqs.\
(\ref{current}) and (\ref{distr}) is determined by the total
resistance,
\begin{equation}
\overline J=\sum_\alpha \int d\epsilon \overline j_\alpha
(\epsilon) = \frac{1}{eR_0} \int d\epsilon ~[f_L-f_R] =
\frac{V}{R_0},
\end{equation}
where the bias $eV$ is the difference in the chemical potentials
of the left and right leads $f_L(\epsilon-eV)=f_R (\epsilon)$. In
the absence of spin-flip scattering $s \to 0$ the resistance of
the parallel (the valve switched 'on') configuration tends to the
$2R+2R_N$ value, while for the antiparallel one (the valve
switched 'off') it diverges. For $s \gg 1$, both resistances tend
to $2R+R_N$.

{\it Shot noise.} To solve Eqs.\ (\ref{equations}) it is
convenient to write the fluctuating part of the distribution
function in the form,
\begin{eqnarray}
\label{solutions} \delta f(x)&=&A+Bx+\frac{e}{\sigma} \int\limits
dx'
G_0 (x,x')\frac{\partial {\cal J}}{\partial x'}\\
\delta
f_s(x)&=&A_s\cosh{({x}/{L_s})}+B_s\sinh{({x}/{L_s})}\nonumber\\&+&
\int\limits  dx' G_{s} (x,x') \left(\frac{e}{\sigma}\frac{\partial
{\cal J}_s }{\partial x'}+\frac{{\cal L }}{D} \right),
\end{eqnarray}
with the help of the Green function vanishing at the interfaces,
\begin{equation}
\label{green} G_{s}(x,x')={L_s} \frac{\sinh{({x_</L_s+s})}
\sinh{({x_>/L_s-s})}} {\sinh{(2s)}},
\end{equation}
with $x_<(x_>)$ standing for the smaller (larger) of the two
coordinates $x,x'$. The function $G_0(x,x')$ is determined from
the same expression (\ref{green}) with $s \to 0$. The coefficients
$A,A_s,B,B_s$ are to be determined from the boundary conditions
(\ref{bound}). It should be pointed out that the distributions in
the leads do not fluctuate $\delta f_L=\delta f_R=0$. The
fluctuation of the total current is determined by the coefficient
$B$ only, according to,
\begin{equation}
\label{currentfluc} \delta J =-\frac{2\sigma}{e} B +\frac{2}{L}
\int \limits  dx ~{\cal J}(x).
\end{equation}
Resolving a set of linear algebraic equations (obtained from the
boundary conditions) with respect to $B$ we find the fluctuation
of the total energy-resolved current,
\begin{eqnarray}
\label{deltaj} \delta J(\epsilon,t) &=&\frac{R}{R_0} [{\cal
I}_L+{\cal I}_R]+\frac{R_N}{R_0L}\sum_\alpha \int  dx~ K_\alpha
(x) {\cal J}_\alpha (x) \nonumber\\ &&+\frac{L_s}{eR_0 D} \int dx
~ M(x) {\cal L}(x),
\end{eqnarray}
where the kernel function $K_\alpha(x)$ depends on the valve
configuration,
\begin{equation} K_\alpha (x)=1-\alpha \left\{  \begin{array}{ll}
\frac{\sinh{(x/L_s)}}{\sinh{s}}, &~~~\text{antiparallel}, \\
\frac{\cosh{(x/L_s)}}{\cosh{s}}, &~~~\text{parallel}.
\end{array} \right.
\end{equation}
The static shot noise power determined as the zero-frequency
transform of the current-current correlation function $S=\int dt~
\langle \delta J(t) \delta J(0) \rangle$ can now be calculated
from Eq.\ (\ref{deltaj}) with the help of the correlation
functions for the Langevin sources,
\begin{eqnarray}
\label{noise} S=e\overline J~ \frac{2R^2}{R^2_0}+ \frac{R_N}{R_0^2
L} \sum_\alpha \int d\epsilon dx~  [ K^2_\alpha(x)
\overline{f_\alpha}(1-\overline{f_\alpha}) \nonumber\\ +M^2(x)
\overline{f_\alpha} (1-\overline{f_{-\alpha}})].
\end{eqnarray}
Substituting the mean distribution functions  (\ref{distr}) into
Eq.\ (\ref{noise}) and evaluating the spatial integrals we obtain
the final expressions for the dimensionless noise-to-current
ratio, $F=S/e\overline{J}$, also known as the Fano factor,
\begin{widetext}
\begin{eqnarray}
F_{\downarrow\uparrow}=\frac{r^2s^2}{2p^2_{\downarrow\uparrow}}+\frac{s+\coth{s}}{2p_{\downarrow\uparrow}}+
\frac{s}{2p^3_{\downarrow\uparrow}}
\left(\frac{s[5-\cosh{(4s)}]+2\sinh{(2s)}}{8s\sinh^4{s}}-\frac{s^2}{3}-s\coth{s}
\right),\\
F_{\downarrow\downarrow}=\frac{r^2s^2}{2p_{\downarrow\downarrow}^2}+\frac{s+\tanh{s}}{2p_{\downarrow\downarrow}}+
\frac{s}{2p_{\downarrow\downarrow}^3}
\left(\frac{s[5-\cosh{(4s)}]-2\sinh{(2s)}}{8s\cosh^4{s}}-\frac{s^2}{3}-s\tanh{s}
\right),
\end{eqnarray}
\end{widetext}
with $p=R_0/R_s$ being the dimensionless total resistance:
$p_{\downarrow\uparrow}= s(r+1)+\coth{s}$ for the antiparallel
configuration and $p_{\downarrow\downarrow}= s(r+1)+\tanh{s}$ for
the parallel configuration. We also introduced the dimensionless
tunneling resistance $r=2R/R_N$.

Figs.\ 2 and 3 illustrate the Fano factor behavior with the
spin-flip intensity $s$ for different values of the contact
resistance $r$ for antiparallel and parallel valve configurations
respectively. Let us first discuss the regime of transparent F-N
interfaces $r=0$. For large spin-flip scattering, $s \to \infty$,
the shot noise approaches the universal value $F=1/3$ independent
of the relative magnetization of the leads. This is obvious since
an injected electron quickly loses its polarization. For
intermediate values, $s>1$, the noise is slightly increased by
spin-flip scattering both for parallel and antiparallel spin valve
configurations. For small spin-flip intensity, $s<1$, the noise
behavior is completely different. In the parallel configuration
the Fano factor is returned to its universal value $1/3$, which is
easy to understand by realizing that electric current is
transferred predominantly by the spin-down states. In the
antiparallel configuration, however, the small-amount of spin-flip
scattering is responsible for the finite conductance itself. The
spin-flip induced fluctuations contribute to the noise comparably
to the disorder-induced fluctuations. The noise power is therefore
{\it enhanced} reaching ultimately the {\it full Poissonian value}
usually reflective of the independent electron transmission, like
in a tunnel junction or a Shottky vacuum diode.

The presence of contacts with the finite resistance $r$ changes
the noise-to-current ratio. For large spin-flip scattering,
\begin{equation*}
F_{\downarrow\uparrow} =F_{\downarrow
\downarrow}=\frac{1}{2(r+1)}+\frac{r^2}{2(r+1)^2}-\frac{1}{6(r+1)^3},
\end{equation*}
the Fano factor is increased monotonously from $F=1/3$ to $F=1/2$
by changing $r$ from zero to infinity. Exactly opposite, however,
happens for antiparallel configuration with low spin-flip
scattering ('off'-state of the valve), $s<1$, where the presence
of contacts actually {\it suppresses} the noise power.

The stochastic diffusion equations presented here allow for the
discussion of the time-dependent problems as well, e.g.\ frequency
dependence of the noise power. Without spin-flip scattering the
noise spectrum is white as a result of the Debye screening
\cite{BlB}. Shot noise in a spin valve is different since
fluctuations of spin density do not require fluctuations of charge
density. Mathematically it is illustrated by the existence of the
new (spin-flip) frequency scale $D/L^2_s$. The calculations would
be similar to those performed for the phononic noise spectrum
\cite{MPB}.

Fruitful discussions with B.\ Halperin, D.\ Davidovic and E.\
Demler are gratefully appreciated. This material is based on work
supported by the NSF under grant PHY-01-17795 and by the Defence
Advanced Research Programs Agency (DARPA) under Award No.
MDA972-01-1-0024.


\begin{thebibliography}{50}
\bibitem{t} This formula holds for $T=0$ only. To
consider the effects of thermal fluctuations one needs a more
general expression for the intrinsic noise of a tunnel barrier,
$\overline{ {\cal I} (\epsilon,t){\cal I}(\epsilon',t') }
=\delta(\epsilon-\epsilon')\delta(t-t') R^{-1} [\overline
n_L(1-\overline n_R)+\overline n_R(1-\overline n_L)]$, where $n_L$
and $n_R$ are the distribution functions at the left and right
side of the barrier.
\newpage

\begin{figure}[h]
\resizebox{.31\textwidth}{!}{\includegraphics{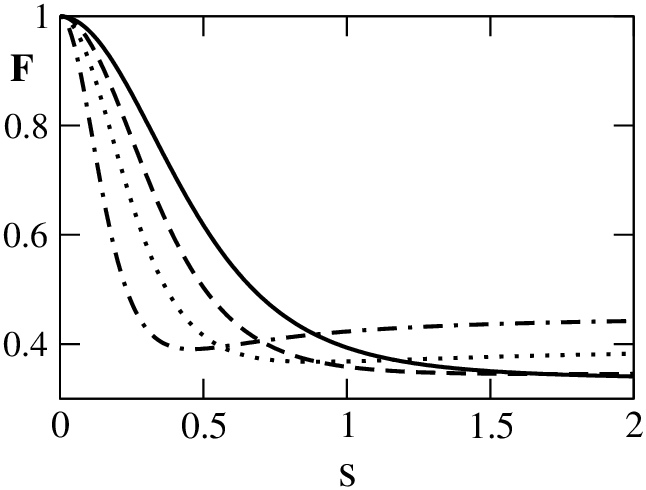}}
\caption{ Fano factor $F_{\downarrow \uparrow}$ versus the
spin-flip intensity  for different values of the contact
resistance: $r=0$ (solid line), $r=1$ (dashed), $r=3$ (dotted),
$r=10$ (dot-dashed).}
\end{figure}
\newpage

\begin{figure}[h]
\resizebox{.32\textwidth}{!}{\includegraphics{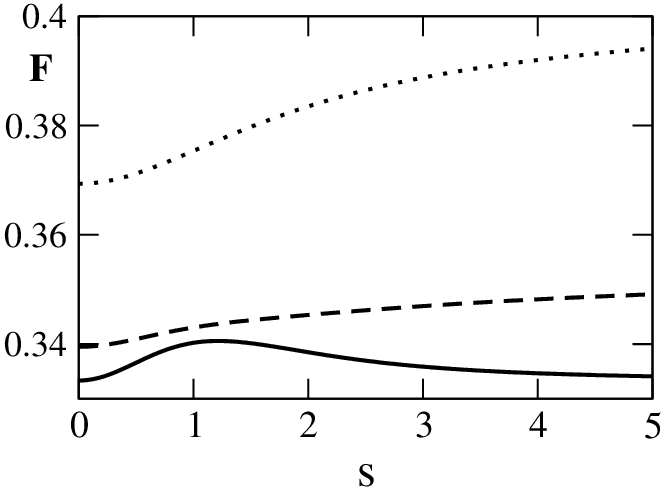}} \caption{
Fano factor $F_{\downarrow \downarrow}$ versus the spin-flip
intensity  for different values of the contact resistance: $r=0$
(solid line), $r=1$ (dashed), $r=3$ (dotted).}
\end{figure}
\end{thebibliography}
\end{document}